\documentclass[make,article,accept,pdftex,moreauthors]{Definitions/mdpi} 
\usepackage[justification=centering]{caption}
\usepackage{multicol}
\usepackage[most]{tcolorbox}
\usepackage{refcount}
\usepackage{enumerate}

\newtcolorbox[auto counter, number within=section]{mybox}[2][]{
  enhanced,
  colback=gray!5,
  colframe=gray!60!white,
  title=Box~\thetcbcounter: #2,
  fonttitle=\scriptsize\bfseries,
  coltitle=black,
  fontupper=\scriptsize,
  boxrule=0.3pt,
  arc=1pt,
  left=2pt,
  right=2pt,
  top=2pt,
  bottom=2pt,
  boxsep=1pt,
  sharp corners,
  #1
}
\firstpage{1} 
\pubvolume{1}
\issuenum{1}
\articlenumber{0}
\pubyear{2025}
\copyrightyear{2025}
\externaleditor{Firstname Lastname} 
\datereceived{26 June 2025} 
\daterevised{19 July 2025} 
\dateaccepted{22 July 2025} 
\datepublished{ } 
\hreflink{https://doi.org/} 

\Title{AlzheimerRAG: Multimodal Retrieval-Augmented Generation for Clinical Use Cases}

\TitleCitation{AlzheimerRAG: Multimodal Retrieval-Augmented Generation for Clinical Use Cases}

\Author{{Aritra Kumar Lahiri} $^{1,}${*}\orcidA{} and {Qinmin Vivian Hu} $^{2}$\orcidB{} }

\AuthorNames{Aritra Kumar Lahiri and Qinmin Vivian Hu}

       \AuthorCitation{Lahiri, A.K.; Hu, Q.V.}

\address{%
$^{1,2}$ \quad Department of Computer Science, Toronto Metropolitan University, Toronto, ON M5B 2K3, Canada}
\corres{Correspondence: aritra.lahiri@torontomu.ca}

\abstract{Recent advancements in generative AI have fostered the development of highly adept Large Language Models (LLMs) that integrate diverse data types to empower decision-making. Among these, multimodal retrieval-augmented generation (RAG) applications are promising because they combine the strengths of information retrieval and generative models, enhancing their utility across various domains, including clinical use cases. This paper introduces AlzheimerRAG, 
 a multimodal RAG application for clinical use cases, primarily focusing on Alzheimer's disease case studies from PubMed articles. This application incorporates cross-modal attention fusion techniques to integrate textual and visual data processing by efficiently indexing and accessing vast amounts of biomedical literature. Our experimental results, compared to benchmarks such as BioASQ and PubMedQA, yield improved performance in the retrieval and synthesis of domain-specific information. We also present a case study using our multimodal RAG in various Alzheimer's clinical scenarios. We infer that AlzheimerRAG can generate responses with accuracy non-inferior to humans and with low rates of hallucination.}


\begin{document}




\section{Introduction}

\label{sec:introduction}
The high volume and variety of data in medical research offer several opportunities and challenges. Of these, Alzheimer's disease (AD) is a particularly compelling case study because it is multicausal, involving genetic, biochemical, and environmental factors, and also involves complex clinical presentations. Despite the tremendous progress, few effective methods exist for diagnosing, treating, and preventing Alzheimer's disease (AD). This knowledge gap is further exacerbated by the growing volume and fragmentation across various data modalities, including textual descriptions, clinical trial data, imaging studies, and molecular data. Traditional methods of synthesizing such a large volume of knowledge are ineffective; most have a single-modality approach, which may miss the insights obtained synergistically from integrated data. This gap in methodology underscores the need for a robust, unified framework that can leverage multiple modalities to enhance the retrieval process by making it more context-aware and reducing the retrieval of irrelevant or less pertinent information.

In this research, we describe a novel multimodal retrieval-augmented generation (RAG) application, \textbf{AlzheimerRAG}
 \href{https://youtu.be/lR2pDjNSaRg}{(video demonstration)}, which integrates textual and visual modalities to improve contextual understanding and information synthesis from the biomedical literature. Our primary research objective in implementing multimodal RAG is to enhance context-aware retrieval capabilities by integrating heterogeneous data types, including textual data, images, and clinical trial information from PubMed articles.  Existing methods~\cite{chen2024benchmarking,zhao2023retrieving,braunschweiler2023evaluating,wang2024biorag} 
 typically, focus on textual or visual data separately, leaving a gap for integrated multimodal solutions. Latest research in Context-Aware Retrieval~\cite{lewis2020retrieval, seonwoo2020contextaware, sticha2023utilizing, dragonverseqa, hodqa, gotqa} provides the foundation for RAG models by demonstrating how retrieval could enhance the generation capabilities of language models, particularly in knowledge-intensive tasks. Integrating RAG methodologies with multimodal inputs is a burgeoning area of research, as highlighted by Xia et al.~\cite{xia2024mmed}, who proposed a multimodal RAG system that enhanced data synthesis across text and image modalities. In light of these advancements, the novelty of our approach lies in the seamless integration and alignment of multimodal data during the cross-modal attention fusion process. The AlzheimerRAG framework combines rapid, accurate retrieval via object stores with specialized language models, enhancing its capability to address the nuances of multimodal information pertinent to Alzheimer’s disease. We utilize an optimized mechanism for fine-tuning by implementing Parameter-Efficient Fine-Tuning (PEFT)~\cite{ding2023parameter} and inducing cross-modal attention fusion to facilitate the synergistic information flow between the text and image models. The fine-tuned models are then incorporated into a multimodal RAG workflow, developed as a 
 web application with a user interface that allows end-users to retrieve context-aware answers from their queries. The target audience of this application includes biomedical researchers for synthesizing Alzheimer's disease literature and identifying disease trends, clinicians to support diagnosis and treatment planning for AD, and healthcare institutions for clinical trial design and support.

Benchmark datasets such as BioASQ~\cite{tsatsaronis2015overview} and PubMedQA~\cite{jin2019pubmedqa} have been instrumental in measuring the effectiveness of multimodal RAG systems~\cite{chen2022murag}. BioASQ, a large-scale biomedical semantic indexing and question-answering (QA) dataset, provides a robust framework for assessing models' retrieval and QA capabilities. Similarly, PubMedQA offers insights into the accuracy of models in handling biomedical queries, making it an essential tool for evaluating AlzheimerRAG’s performance against existing benchmarks. In comparative studies, models that integrate multimodal data have been shown to outperform traditional single-modality systems. For instance, models like T5~\cite{raffel2020exploring} have been evaluated in the context of biomedical question answering, demonstrating significant gains when multimodal inputs are utilized. This trend reinforces the need for AlzheimerRAG’s multimodal framework to enhance the understanding and treatment of AD.

In summary, our research contributions advance the multimodal RAG domain in AD in the following aspects: 

\begin{itemize}
    \item Context-aware retrieval-augmented generation---Our framework enhances traditional RAG models by prioritizing the context relevance of domain-specific information, thereby increasing accuracy and utility in biomedical applications. 
    \item Advanced cross-modal attention fusion---AlzheimerRAG integrates multimodal data more effectively using transformer architectures and cross-modal attention mechanisms tailored to handle heterogeneous data types.
    \item RAG user interface---Our system implements multimodal RAG as a web-based application using the latest state-of-the-art technologies like LangChain, FastAPI, Jinja2, and FaissDB to provide users with a robust interface for performing biomedical information retrieval tasks through the context-aware question-answering paradigm.
    \item Comparable framework with state-of-the-art benchmarks---We evaluate the capability of the Multimodal RAG application with benchmark datasets like BioASQ and PubMedQA, along with other comparable LLM RAG models. We also study the effectiveness of our AlzheimerRAG against human-generated responses for different clinical scenarios in Alzheimer's disease to gauge the accuracy and hallucination rates of the retrieved answers.
\end{itemize}

Although 
 this research is rooted in the Alzheimer's domain, the usage of cross-modal attention fusion in multimodal RAG makes it adaptable to any domain that requires alignment of textual data with visuals, audio, or structured data. The modular approach of AlzheimerRAG supports adapting it to answering capabilities on queries related to comparable medical domains linked to Alzheimer’s Disease. Additionally, the technical scalability is also enhanced by parameter-efficient fine-tuning techniques such as QLoRA, which enables efficient fine-tuning of LLMs (e.g., LlaMA and LLaVA) on niche datasets without full retraining.  

\section{Related Work}
The AlzheimerRAG framework was developed within the rapidly evolving landscape of multimodal data integration and retrieval-augmented generation techniques, which are becoming increasingly crucial in biomedical research. Recent studies have demonstrated the importance of leveraging multiple data modalities to enhance diagnosis, treatment, and understanding of complex diseases, such as Alzheimer's. 

Existing research~\cite{fang2024gfe, yang2023large, yang2023integrating,bolton2024biomedlm,treder2024introduction} has highlighted the efficacy of attention mechanisms that span multiple modalities, which are instrumental in synthesizing heterogeneous information sources in medical contexts. For example, the effectiveness of multimodal token fusion for vision transformers~\cite{wang2022multimodal} has been demonstrated, which significantly improves the integration of visual and textual data in medical imaging~\cite{jiang2023deep}. Similarly, cross-modal translation and alignment techniques~\cite{zhou2023cross, zhang2022robust} have been showcased that facilitate survival analysis, emphasizing the benefits of integrating diverse data types to yield richer insights. Additionally, recent developments in knowledge distillation have further enhanced model efficiency in healthcare applications, as demonstrated in the work by Hinton et al. and Gupta et al.~\cite{hinton2015distilling, gupta2023packd}, which involves transferring knowledge from a larger model (teacher) to a smaller model (student), thereby retaining performance while reducing computational costs. Various studies have adopted this methodology, notably, the work that discovered integrating imaging and genetic data improved predictive outputs in Alzheimer’s models~\cite{zhao2023retrieving}.

The application of AI in Alzheimer’s research has been underscored by studies~\cite{liu2018use} which leverage multimodal inputs to improve early diagnosis and patient stratification. Other research, such as~\cite{li2024studying, yao2023artificial}, has focused on using AI to manage Alzheimer’s disease symptoms, demonstrating that AI-driven solutions can provide valuable insights and recommendations for patient care. The BioBERT model~\cite{lee2020biobert} represents a significant advancement in biomedical text mining, emphasizing the utility of transformer models fine-tuned for biomedical applications. This model has been foundational in developing various biomedical applications, including those focused on Alzheimer’s disease, where precision in information retrieval is critical. RAG methodologies~\cite{madan2024transformer} have gained traction in biomedical research for efficiently synthesizing information from large datasets. The works~\cite{lewis2020retrieval,xiong2024benchmarking} laid the ground for RAG models by demonstrating how retrieval could enhance the generation capabilities of language models, particularly in knowledge-intensive tasks. This has profound implications for healthcare, where accurate and timely information retrieval can guide clinical decisions.

Compared to these advancements, the AlzheimerRAG framework combines rapid, accurate retrieval via FaissDB with specialized language models, enhancing its capability to address the nuances of multimodal information pertinent to Alzheimer’s Disease.

\section{Materials and Methods}

\begin{figure}[H]
    \includegraphics[width=\textwidth]{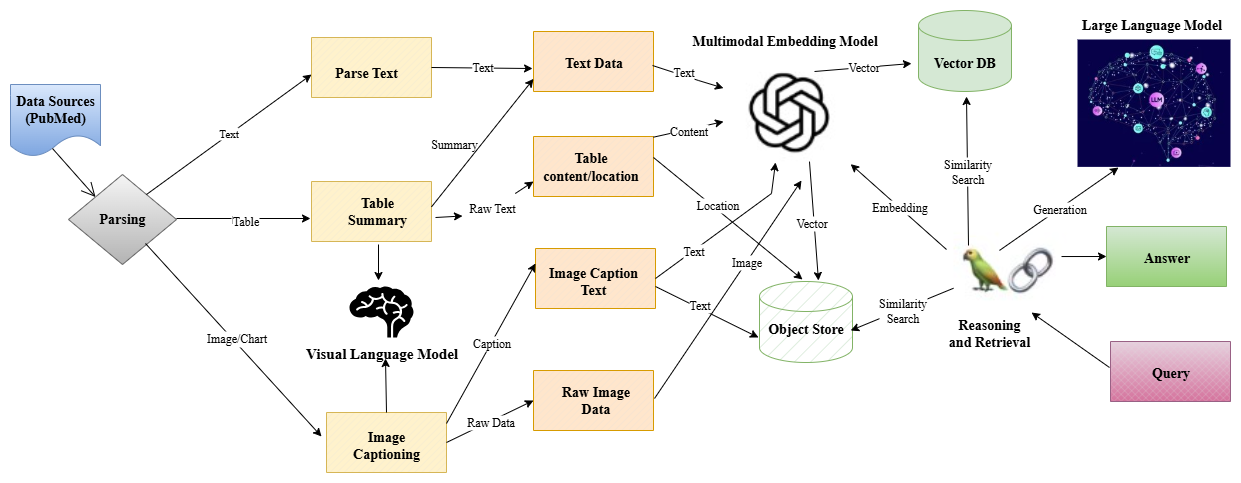}
    \caption{AlzheimerRAG 
 architecture.} 
    \label{fig1}
\end{figure}

The overall architecture of AlzheimerRAG is described in Figure \ref{fig1}. To simplify, the architecture diagram in Figure \ref{fig1} illustrates a multimodal RAG  component for the biomedical literature (e.g., PubMed), where text, tables, and images are extracted through parsing and processed separately—text is parsed, tables are summarized, and images are captioned using a visual language model. These processed elements are then converted into embeddings through a cross-modal embedding fusion method and stored in an object store and a vector database. Upon receiving a user query, the system retrieves relevant information using similarity search and passes it to a large language model, which generates a context-aware answer by reasoning over the retrieved multimodal content.

In the subsequent sections, we describe each step in the architecture flow in more detail, followed by a demonstration of the application with the technical components.

\vspace{-3pt}

\subsection{Data Collection and Preprocessing}
The first step of our process involved collecting relevant articles from PubMed. We accomplished this by writing a Python script~\cite{lahiri2024tmu} that called the National Centre for Biotechnology Information (NCBI) Entrez Programming Utilities (E-utilities) API to fetch the top \mbox{2000~articles} from the PubMed repository~\cite{pubmed} related to the "Alzheimer's Disease" search term~\cite{thomo2024pubmed}. The articles were fetched in batches per API request, adhering to NCBI API rate limits and sorted by relevance during the retrieval process. We parsed each document, collecting the full texts, abstracts, tables, and figures for textual and image retrieval. After that, we cleaned and normalized the data for the data preprocessing step to ensure consistency and usability. This involved removing hyperlinks, references, and footnotes. We also standardized the figures/diagrams format by converting them to a consistent image format for uniform processing.
\subsection{Textual Data Retrieval}
\label{subsec:text_retrieval}
This step retrieves the clinical text data related to Alzheimer's disease (AD) for textual and tabular data processing. In our workflow, for generating the text embedding, we fine-tuned the "Llama-2-7b-pubmed" ~\cite{llama-meta-ai, llama2pubmed} model by training it with the PubMedQA~\cite{jin2019pubmedqa} dataset from HuggingFace. The fine-tuning used parameter-efficient fine-tuning (PEFT) techniques like QLoRA~\cite{dettmers2024qlora}. Table \ref{tab:hyperparam} outlines the QLoRA parameters and the training argument parameters used for fine-tuning. 
\begin{table}[H]

    \caption{QLoRA 
 hyperparameters: LlaMA.}

        \begin{tabularx}{\textwidth}{CC}
        \toprule
        \textbf{Parameter} & \textbf{Value} \\ \midrule
        \multicolumn{2}{c}{\textbf{QLoRA Parameters}} 
 \\ \midrule
        LoRA attention dimension & 64 \\ \midrule
        Alpha parameter for LoRA scaling & 16 \\ \midrule
        Dropout probability for LoRA layers & 0.1 \\ \midrule
        \multicolumn{2}{c}{\textbf{Training Hyperparameters}} \\ \midrule
        Number of training epochs & 1 \\ \midrule
        FP16/BF16 training & False (True for A100 GPU) \\ \midrule
        Training batch size per GPU & 4 \\ \midrule
        Evaluation batch size per GPU & 4 \\ \midrule
        Gradient accumulation steps & 1 \\ \midrule
        Enable gradient checkpointing & True \\ \midrule
        Max gradient norm (clipping) & 0.3 \\ \midrule
        Initial learning rate & 2 $\times$ 10$^{-4}$ 
 \\ \midrule
        Weight decay & 0.001 \\ \midrule
        Optimizer & \texttt{paged\_adamw\_32bit} 
 \\ \midrule
        Learning rate scheduler & \texttt{cosine} \\ \midrule
        Number of training steps & $-$1 \\ \midrule
        Warmup ratio & 0.03 \\ \bottomrule
        \end{tabularx}%

    \label{tab:hyperparam}
\end{table}

\subsubsection*{\textbf{Textual and Tabular Data Processing}} 
 The extracted data were chunked into structured text and table summaries.  Then, a layout model (for tables) and titles were used for candidate sub-sections of the document (e.g., Introduction, Methods, etc.). Finally, post-processing was conducted to aggregate text under each title, and further chunking into text blocks was performed for downstream processing based on user-specific flags for each block. After that step, the text embeddings converted the smaller blocks into embedding vectors, which were used for cross-modal attention fusion.

\subsection{Image Retrieval}
\label{subsec:image_ret}
For the generation of feature embeddings that capture image details from the PubMed articles, we fine-tuned the "LlaVA" (Language and Vision Assistant Model, version 2)~\cite{zhu2024llava} model using the official LLaVA repo with the Llama-2 7B backbone language model~\cite{llamaindex}. LLaVA combines pre-trained language models (such as Vicuna or LLaMA~\cite{touvron2023llama,roziere2023code}) with visual models (such as CLIP's~\cite{radford2021learning} visual encoder) by converting visual features into embeddings that are compatible with the language model. Its training has two stages: a pre-training stage, where image--text pairs align visual and language embeddings with only the projection matrix being trained~\cite{marino2019ok}, and a fine-tuning stage, where the visual encoder remains frozen while the projection layer and language model are updated~\cite{xia2024rulereliablemultimodalrag}. Using the fine-tuned approach preserves the strengths of the large language model while lowering computational requirements, making it ideal for resource-limited environments and quick adaptation to new data. The hyperparameters used for fine-tuning are presented in Table \ref{tab:llava}. QLoRA uses the 4-bit NormalFloat, which is explicitly designed for customarily distributed weights, thereby further reducing memory usage.
\begin{table}[H]
        \caption{LlaVA hyperparameters} 
        \begin{tabularx}{\textwidth}{CC}
        \toprule
        \textbf{Parameter} & \textbf{Value} \\ \midrule       
        {lora\_enable} & {True} \\ \midrule
        {lora\_r} & {128} \\\midrule
        {lora\_alpha} & {256} \\\midrule
        {mm\_projector\_lr} & {2 $\times$ 10$^{-5}$} \\\midrule
        {bits} & {4} \\\midrule
        {learning\_rate} & {2 $\times$ 10$^{-4}$} \\\midrule
        {weight\_decay} & {0.001} \\\midrule
        {warmup\_ratio} & {0.03} \\
\bottomrule
        \end{tabularx}
        \label{tab:llava}
\end{table}

\subsection{Cross-Modal Attention Fusion} 
\begin{figure}[H]
    \includegraphics[width=\textwidth]{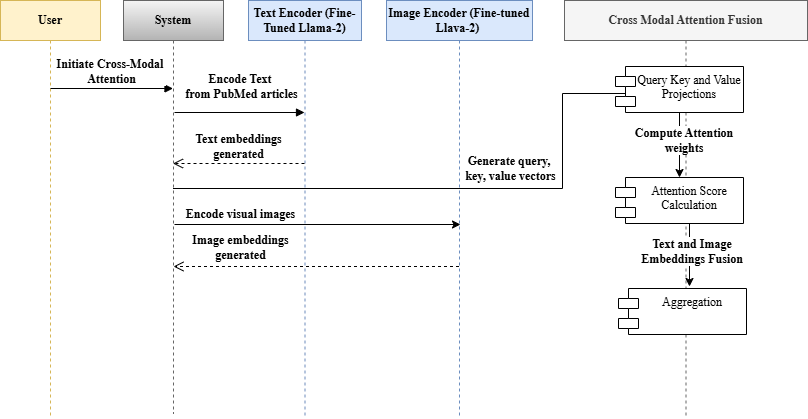}
    \caption{AlzheimerRAG: 
 cross-modal attention sequence diagram.} 
    \label{fig3}
\end{figure}

Cross-modal attention fusion is a mechanism that facilitates interaction between different modalities, within our current scope, specifically between text and images. It allows a model to selectively focus on relevant parts of both modalities by computing attention weights. These weights are used to modulate the embeddings from each modality, enabling a richer and more comprehensive representation. In our context, the cross-modal attention fusion ensures that the integrated textual and visual data contribute meaningfully to medical information retrieval. The process steps of cross-modal attention fusion are detailed in Figure \ref{fig3} as a sequence diagram. 

The three steps associated with this process are described below: 

\begin{itemize}
    \item Generate query, key, and value vectors from the text and image embeddings from Sections \ref{subsec:text_retrieval} and \ref{subsec:image_ret}, respectively.
    \item Compute the attention scores using the dot-product attention mechanism shown below: 
            \begin{equation}
                \text{scores} = \frac{\text{queries} \cdot \text{keys}^\top}{\sqrt{d_k}}
            \end{equation} 
        where
            \begin{itemize}
                \item \(\text{queries}\) and \(\text{keys}\) are matrices of size \((n \times d_k)\), with \(n\) being the number of tokens and \(d_k\) the dimension of each key.
                \item \(d_k\) is the dimensionality of the keys used for scaling.
                \item \(\sqrt{d_k}\) scales the dot-product, helping to stabilize gradients in deeper networks.
            \end{itemize}
    \item Aggregate contributions from both modalities based on attention weights:
            \begin{equation}
                \text{aggr\_embeddings} = \text{attn\_wts} \cdot \text{values}
            \end{equation}
            where
            \begin{itemize}
                \item \(\text{attn\_wts}\) is a matrix representing the attention scores, with dimensions \((n \times m)\), where \(n\) is the number of tokens, and \(m\) is the dimensionality of each value.
                \item \(\text{values}\) is a matrix of values corresponding to tokens, typically with dimensions \((m \times d)\), where \(d\) is the embedding size.
            \end{itemize}
            The resulting \(\text{aggr\_embeddings}\) is a combination of the values weighted by attention.
            \begin{equation}
                \text{combined\_features} = \text{aggr\_attn}(\text{values}, \text{attn\_wts})
            \end{equation}
\end{itemize}

Finally, the combined feature embeddings are indexed as vectors in an object store, which allows quicker retrieval of multimodal data. 

\subsection{AlzheimerRAG Demonstration}
\subsubsection{System Walk-Through}
AlzheimerRAG is implemented as a Python (version 3.8.x) Web Application utilizing FastAPI and Jinja2 Templates with LangChain integration. It provides a simple user interface \href{https://pubmed-multimodal-rag-ae786f93140b.herokuapp.com/}{(Web Application)}  for leveraging efficient multimodal RAG capabilities related to AD. The application 
\href{https://tinyurl.com/AlzheimerRAG}{(Source Code)}
is deployed in Heroku, a cloud-based Platform-as-a-Service (PaaS) solution that helps manage seamless continuous integration and deployment. It provides the functionality for information retrieval from user queries. The multimodal RAG component extracts context-aware relevant images as part of the output response. The demo video can be accessed from this link \href{https://youtu.be/lR2pDjNSaRg}{(Video Demonstration)} 

\begin{figure}[H]
    \includegraphics[width=\textwidth]{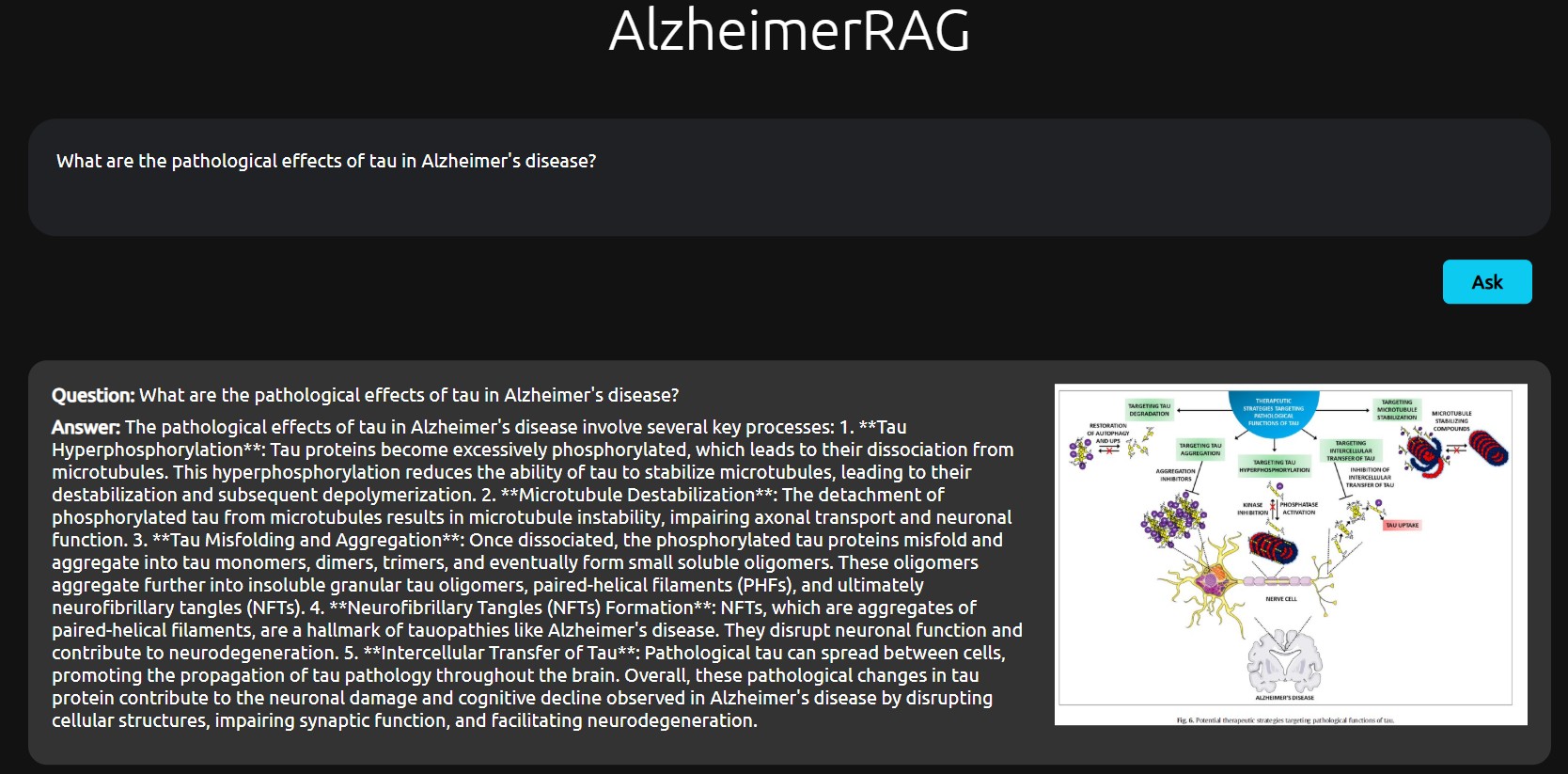}
    \caption{AlzheimerRAG 
: sample user interface response.} 
    \label{fig:alzheimerrag2}
\end{figure}

A sample response from the AlzheimerRAG user interface can be observed in Figure~\ref{fig:alzheimerrag2}, where relevant text and images are fetched for a particular user query related to Alzheimer's disease from the embedded PubMed articles.

\subsubsection{Key Technical Components}
The key technical components are summarized below. 

FastAPI for API development: 
 FastAPI is a high-performance web framework for API development that provides an intuitive interface for API development and integrates seamlessly with Python’s async capabilities.

Jinja2 for template rendering: Jinja2 is a templating engine for Python that offers dynamic template rendering. It serves HTML content from backend data, enabling a seamless and interactive user experience.

FaissDB for embedding multimodal data: FaissDB~\cite{meta_faiss}, a vector DB, is widely used for embedding multimodal data. Embedding is the process of converting content into a numerical representation (i.e., vectors) for large language learning models and is crucial for transforming preprocessed healthcare knowledge into individual vectors. The text and image embeddings are encoded into uniform, high-dimensional vectors and indexed for efficient similarity searches. When a query is made, the reasoning and retrieval component searches the vector space to extract relevant information. The benefit is that it uses an approximate nearest neighbor (ANN) search to quickly locate embeddings in high-dimensional space, which is essential for large-scale applications. The generation component uses the retrieved multimodal representations to produce outputs in various formats, such as text or images.

LangChain as a Retrieval Agent for multimodal RAG: The Retrieval Agent is a medium to pinpoint the most relevant knowledge in response to user queries. This process involves using the embedding model to convert the user query text into vectors, which are then searched through the vector storage to identify the closest matching vectors. The effectiveness of a Retrieval Agent is closely tied to the underlying framework upon which it is built. Therefore, we utilized the LangChain~\cite{langchain2021,topsakal2023creating} framework, a premium existing open-source framework, along with LlamaIndex~\cite{llamaindex} because of its significant advantages in \mbox{(i) preservation} of table data integrity, \mbox{(ii) streamlining} the handling of Multimodal data, \mbox{(iii) enhanced} semantic embedding. Together, LlamaIndex and LangChain enhance the context-awareness of extracted content, enabling efficient retrieval and synthesis of information and producing nuanced outputs.

\section{Experimental Results}
\subsection{Comparative Evaluation}
We compared our AlzheimerRAG application against state-of-the-art techniques in the biomedical domain and evaluated the performance of our methods. In our experiments, we selected BioBERT~\cite{lee2020biobert}, a transformer model fine-tuned on biomedical text, and MedPix~\cite{datadiscovery}, which utilizes deep learning for medical image classification. To compare the cross-modal attention fusion, we introduced a naive fusion of text and image modalities among two models, primarily by concatenating the embeddings without significant interaction between the modalities. Among the newer variants, we included PubMedBERT~\cite{gu2024pubmedbert}, LlaVA-Med~\cite{li2024llava}, and BioRAG~\cite{wang2024biorag} in our evaluation. 

Table \ref{table:performance_comparison} represents the performance, where it is observed that AlzheimerRAG, with its multimodal RAG design, retained the lead over LlaVA-Med, a multimodal model for biomedicine that lacks retrieval capabilities, and BioRAG, a text-only RAG model with PubMed integration.


\begin{table}[H]
    \caption{AlzheimerRAG evaluation with comparative benchmark models.}

    \begin{tabularx}{\textwidth}{CCCC}
\toprule
        \textbf{Model} & \textbf{Recall} & \textbf{Precision@10} & \textbf{F1} \\ \midrule
        BioBERT & 0.72 & 0.69 & 0.71  \\ \midrule
        MedPix & 0.65 & 0.62 & 0.63  \\ \midrule
        BioBERT + MedPix & 0.78 & 0.75 & 0.76  \\ \midrule
        PubMedBERT & 0.80 & 0.77 & 0.78  \\ \midrule
        LlaVA-Med & 0.82 & 0.79 & 0.80  \\ \midrule
        BioRAG & 0.87 & 0.84 & 0.85 \\ \midrule
        AlzheimerRAG 
 & 0.88 & 0.85 & 0.86 \\ \bottomrule
    \end{tabularx}
    \label{table:performance_comparison}
\end{table}
Against benchmark datasets like BioASQ~\cite{tsatsaronis2015overview}, a large-scale biomedical semantic indexing and question-answering dataset, and PubMedQA~\cite{jin2019pubmedqa}, developed for QA tasks using a PubMed corpus, we assessed the capability of our multimodal RAG by evaluating the document retrieval from given queries and generating accurate answers to Alzheimer-related questions from the data against GPT-4. The results are highlighted in Table \ref{table:benchmark_comparison}.


\begin{table}[H]
        \caption{Benchmark dataset evaluation: AlzheimerRAG vs. GPT-4.}

            \begin{tabularx}{\textwidth}{CCCC}
\toprule
                \textbf{Benchmark} & \textbf{Metrics} & \textbf{AlzheimerRAG} & \textbf{GPT-4} \\
                \midrule
                \multirow{5}{*}{\textbf{BioASQ}} 
                    & \textbf{Precision@10} & 0.71 & 0.70 \\
                    & \textbf{Recall}       & 0.80 & 0.78 \\
                    & \textbf{MAP}          & 0.78 & 0.74 \\
                    & \textbf{QA Accuracy}  & 0.72 & 0.76 \\
                    & \textbf{F1-Score}     & 0.75 & 0.77 \\
                \midrule
                \multirow{3}{*}{\textbf{PubMedQA}} 
                    & \textbf{Accuracy}     & 0.74 & 0.78 \\
                    & \textbf{Exact Match}  & 0.71 & 0.73 \\
                    & \textbf{F1-Score}     & 0.76 & 0.79 \\
\bottomrule
            \end{tabularx}
        \label{table:benchmark_comparison}
\end{table}

The metrics used in our evaluation included (i) Precision@k, which measures the relevance of the top-k(10) retrieved document; (ii) Recall, which evaluates how many relevant documents are retrieved from the corpus; and (iii) Mean Average Precision (MAP), which provides the mean average precision values for all queries. In terms of question-answering tasks, our evaluation metrics included accuracy (percentage of correctly answered questions), Exact Match (EM) (percentage of questions that are responded to with exact word matches to the ground truth), and F1-score (considers both precision and recall for evaluating answer span quality).  
We further conducted a comparative qualitative evaluation with other models adaptable for the biomedical domain, focusing on retrieval and question-answering capabilities, as depicted in Table \ref{table:model_comparison}. The comparison results are presented by considering the GLUE (General Language Understanding Evaluation)~\cite{wang2019glue} and SuperGLUE (Super General Language Understanding Evaluation)~\cite{wang2020supergluestickierbenchmarkgeneralpurpose} benchmarking leaderboards, which serve as metrics for evaluating how well NLP models handle a wide range of complex and straightforward natural language understanding tasks. 
It can be observed that BioBERT~\cite{lee2020biobert} stands out in biomedical applications due to its PubMed pre-training, achieving high precision in retrieval. SciBERT~\cite{beltagy2019scibert}, with its broader scientific text pre-training, is more versatile but may need fine-tuning for top biomedical QA tasks. BM25~\cite{robertson2009probabilistic}, as a traditional keyword-based model, sets a baseline but lacks deep semantic understanding. ColBERT~\cite{khattab2020colbert} combines efficient retrieval with semantic depth, though it performs moderately without specific domain adjustments. The BERT+TF-IDF~\cite{devlin2018bert} hybrid model strikes a balance between deep learning and traditional retrieval, yielding reasonable results but limited contextual depth. Lastly, T5~\cite{raffel2020exploring} excels in QA, especially when fine-tuned for biomedical contexts, leveraging its generative capabilities to achieve high accuracy. In comparison to these, AlzheimerRAG combines fast, accurate retrieval via FaissDB with specialized language models, making it a powerful tool for biomedical retrieval and QA. Its ability to handle text and images offers a significant advantage in contexts where visual data is essential~\cite{liang2021multibench}.
\begin{table}[H]
\caption{Comparison of benchmark models across GLUE and SuperGLUE metrics.}

\begin{adjustwidth}{-\extralength}{0cm}
\begin{tabularx}{\fulllength}{ccccC}
\toprule
\textbf{Model} & \textbf{BioASQ (Retrieval)} & \textbf{PubMedQA (QA)} & \textbf{Domain} & \textbf{Multimodal}  \\
\midrule
AlzheimerRAG  
 & High precision and recall & High accuracy and F1 & Biomedical & Yes \\
\midrule
BioBERT & High for text & Good accuracy & Biomedical & No \\
\midrule
SciBERT & High for scientific texts & Moderate, versatile & Scientific & No \\
\midrule
BM25 (Baseline) & Fair, keyword-based & Basic QA & N/A & No \\
\midrule
ColBERT & Efficient & Moderate & General-purpose & No \\
\midrule
BERT+TF-IDF (for QA) & Fair & Moderate & General-purpose & No \\
\midrule
T5 (fine-tuned) & Good, versatile & High in QA when fine-tuned & General-purpose & Emerging \\
\bottomrule
\end{tabularx}%
\end{adjustwidth}

\label{table:model_comparison}
\end{table}


\subsection{Ablation Studies}
The primary objective of our ablation studies was to assess the significance of critical components in our mechanism. We conducted multiple combinations for our experiments by removing the cross-modal attention mechanism, QLoRA fine-tuning techniques, and multimodal integration. Each of these simulations was designed to isolate and evaluate the impact of the specific component. 

By removing cross-modal attention, 
 we anticipated that the model's ability to integrate and leverage text and image data effectively would degrade. We replaced the cross-modal attention mechanism with a simple text and image embedding concatenation. Similarly, we fine-tuned the techniques without QLoRA to observe the computation costs and performance. Lastly, we removed the multimodal integration to check whether the model's overall performance would decrease. 

Each variation's performance metrics were recorded and consolidated in Table \ref{table:ablation}.
\begin{table}[H]
\caption{Ablation studies across multiple components.}

\begin{tabularx}{\textwidth}{cCCC}
\toprule
\textbf{Experiment} & \textbf{Recall} & \textbf{Precision} & \textbf{F1 Score}  \\
\midrule
Baseline (AlzheimerRAG) & 0.88 & 0.85 & 0.86  \\\midrule
Without Cross-Modal Attention & 0.75 & 0.72 & 0.74 \\\midrule
Without QLora Fine-Tuning & 0.80 & 0.77 & 0.78 \\\midrule
Without Multimodal Integration & 0.70 & 0.68 & 0.69  \\
\bottomrule
\end{tabularx}
\label{table:ablation}
\end{table}

As observed, Cross-modal attention enables effective interaction between text and image data, with its removal leading to considerable metric degradation. QLoRA fine-tuning improves precision and clinical relevance with lower computational costs than traditional methods. Lastly, multimodal integration is essential to the framework's overall effectiveness, as isolating text and image processing substantially reduces recall, precision, and practical application.

\section{Clinical Case Study Analysis}
We designed a case study to evaluate AlzheimerRAG in clinical scenarios related to AD using five primary clinical scenarios---(1) \textbf{Early Diagnosis and Monitoring}, (2) \textbf{Medication Management}, (3) \textbf{Non-Pharmacological Interventions}, (4) \textbf{Caregiver Support and Education}, (5) \textbf{Behavioral Symptom Management}. The clinical scenario descriptions are provided in Box \ref{box1}.

\begin{boxenv}[H]
\caption{Clinical scenarios.}
\fbox{\small\parbox{\columnwidth-2\fboxsep}{
\begin{itemize}
    \item \textbf{Early Diagnosis and Monitoring}: Assess the system's ability to recommend diagnostic tools and interpret results for early detection.
    \item \textbf{Medication Management}: Determine the ability to guide current medications, potential side effects, and interactions specific to Alzheimer's treatments.
    \item \textbf{Non-Pharmacological Interventions}: Evaluate recommendations for cognitive therapies, physical activities, and lifestyle modifications to slow disease progression.
    \item \textbf{Caregiver Support and Education}: Assess the capability to generate materials for educating caregivers about disease progression and management strategies.
    \item \textbf{Behavioral Symptom Management}: Evaluate the effectiveness of offering strategies to manage common symptoms like agitation, depression, and anxiety.
\end{itemize}}}
\label{box1}
\end{boxenv}

\subsection{System Evaluation}
The clinical scenarios were identified from the medical literature~\cite{atri2019alzheimer, murray2011neuropathologically, campillo2016modelling, amugongo2024retrieval, petersen2010alzheimer, scheltens2021alzheimer, weller2018current, lane2018alzheimer, twarowski2023inflammatory, rostagno2022pathogenesis, eratne2018alzheimer, mantzavinos2017biomarkers, ogbodo2022alzheimer, huang2020clinical, knapskog2021alzheimers, oboudiyat2013alzheimer, aisen2017path, mangialasche2010alzheimer} due to their recognized importance in Alzheimer's treatment. A total of 350 responses were evaluated, comprising 50 human-generated, 150 LLM-generated, and 150 LLM-RAG-generated responses. The correctness of the responses was determined by simulating established guidelines~\cite{ke2024development} and expert reviews~\cite{mturk}. The validation criteria were factual correctness, absence of hallucinations, and clinical applicability.

\textbf{Selection of Domain Experts.} 
 The domain experts selected for the study were senior researchers from the Vector Institute specializing in the biomedical domain and with a strong familiarity with PubMed literature.
 
\begin{figure}[H]
\includegraphics[width=\textwidth]{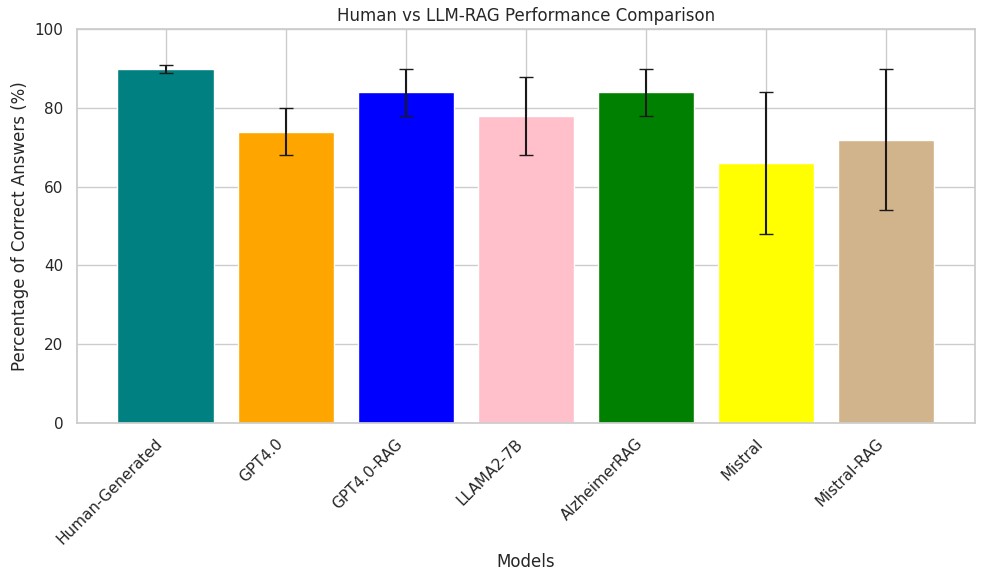}
\caption{Percentage of correct answers: LLM and LLM-RAG groups.} 
\label{fig:llmrag}
\end{figure}

Human-generated answers, provided by domain experts described in Section \ref{subsec:rag_responses}, were used as a comparison. Figure \ref{fig:llmrag} represents the LLM-RAG performance regarding correct-answer percentages. 

Evaluation criteria concerned accuracy and safety. Responses with at least 75\% accuracy in instructions were deemed "correct". However, any response containing a significant medical error was categorized as "wrong (hallucination)". Table \ref{tab:correctresponse} 
indicates the accuracy and the hallucination rate results; AlzheimerRAG (84\%) and GPT4.0-RAG (84\%) were the best-performing RAGs compared to the human-generated answers. 


\begin{table}[H]
\caption{Accuracy and hallucination response: human-generated vs. LLM and LLM-RAG.}
\tablesize{\fontsize{8}{8}\selectfont}

\begin{adjustwidth}{-\extralength}{0cm}
\begin{tabularx}{\fulllength}{m{1.88cm}<{\centering}m{1.88cm}<{\centering}m{1.88cm}<{\centering}m{1.88cm}<{\centering}m{1.88cm}<{\centering}m{1.88cm}<{\centering}m{1.88cm}<{\centering}m{2cm}<{\centering}}
\toprule
\textbf{Models} & 
\textbf{Early Diagnosis  and Monitoring} & 
\textbf{Medication Management} & 
\textbf{Non-Pharma  Interventions }& 
\textbf{Caregiver-Support  and Education }&  
\textbf{Behavioral  Symptom  Management} & 
\textbf{Total  Correct} & 
\textbf{Hallucinations  Present}   \\
\midrule

Human-Generated & 
10/10  (100.0\%) & 
8/10 (80.0\%) & 
9/10 (90.0\%) & 
9/10 (90.0\%) & 
9/10 (90.0\%) & 
45/50 (90.0\%) & 
 -  \\
\midrule

GPT4.0 & 
9/10 (90.0\%) & 
9/10 (90.0\%) & 
6/10 (60.0\%) & 
7/10 (70.0\%) & 
6/10 (60.0\%) & 
37/50 (74.0\%) & 
(3/50) 6\% \\ 
\midrule

GPT4.0-RAG & 
10/10 (100.0\%) & 
10/10 (100.0\%) & 
8/10 (80.0\%) & 
8/10 (80.0\%) & 
6/10 (60.0\%) & 
42/50 (84.0\%) & 
(3/50) 6\% \\ 
\midrule

LLAMA2-7B & 
9/10 (90.0\%) & 
9/10 (90.0\%) & 
7/10 (70.0\%) & 
7/10 (70.0\%) & 
7/10 (70.0\%) & 
39/50 (78.0\%) & 
(5/50) 10\% \\ 
\midrule

AlzheimerRAG & 
10/10 (100.0\%) & 
9/10 (90.0\%) & 
7/10 (70.0\%) & 
8/10 (80.0\%) & 
8/10 (80.0\%) & 
42/50 (84.0\%) & 
(3/50) 6\% \\ 
\midrule

Mistral & 
8/10 (80.0\%) & 
8/10 (80.0\%) & 
5/10 (50.0\%) & 
6/10 (60.0\%) & 
6/10 (60.0\%) & 
33/50 (66.0\%) & 
(9/50) 18\%  \\
\midrule

Mistral-RAG & 
8/10 (80.0\%) & 
8/10 (80.0\%) & 
6/10 (60.0\%) & 
7/10 (70.0\%) & 
7/10 (70.0\%) & 
36/50 (72.0\%) & 
(9/50) 18\% \\ 
\bottomrule

\end{tabularx}
\end{adjustwidth}
\label{tab:correctresponse}
\end{table}

\subsection{Statistical Evaluation}
We also used statistical tools, such as Cohen's H and the chi-square test, to evaluate and compare the performance of human-generated responses against the AlzheimerRAG responses~\cite{efron1994introduction, student1908probable}.

Cohen's H~\cite{cohen2013statistical} is a measure for evaluating the effect size of differences between two proportions. Since the number of answers obtained in our experimental evaluation differed for human-generated and Alzheimer's responses, this metric can provide us with context on the accuracy of the responses between them.

The chi-square test~\cite{Pearson1900,neyman1933ix} can assess whether there is a significant association or difference in the responses generated between the two categories. It is helpful to test the differences in the distribution of responses across multiple clinical scenarios. 

Table \ref{tab:stateval} provides the statistical evaluation results of the different clinical scenarios used in our analysis. In the case of Early Diagnosis and Monitoring, both proportions were the same, inferring there was no difference; hence, the chi-square value was zero. For other clinical scenarios, the results indicated small effect sizes, with notable differences observed only in the Medication Management category. Thus, from the overall results, it can be concluded that there were no major statistically significant differences between the human-generated and the AlzheimerRAG answers.

\begin{table}[H]
    \caption{Comparison between human and AlzheimerRAG answers.}
\tablesize{\fontsize{10}{10}\selectfont}
    \begin{tabularx}{\textwidth}{cCC}
        \toprule
        \textbf{Clinical Scenarios} & \textbf{Cohen's h} & \textbf{Chi-Square} \\ \midrule
        Early Diagnosis and Monitoring & 0     & 0     \\ \midrule
        Medication Management &  $-$0.234     & 0.3922  \\\midrule
        Non-Pharma Interventions & 0.404 & 1.25  \\ \midrule
        Caregiver-Support and Education & 0.234 & 0.3922 \\ \midrule
        Behavioral Symptom Management & 0.234 & 0.3922 \\ \bottomrule
    \end{tabularx}
    \label{tab:stateval}
\end{table}
\clearpage 
\subsection{AlzheimerRAG Responses vs. Human-Generated Clinical Scenario Responses}
\label{subsec:rag_responses}
\begin{figure}[H]

\includegraphics[width=\textwidth]{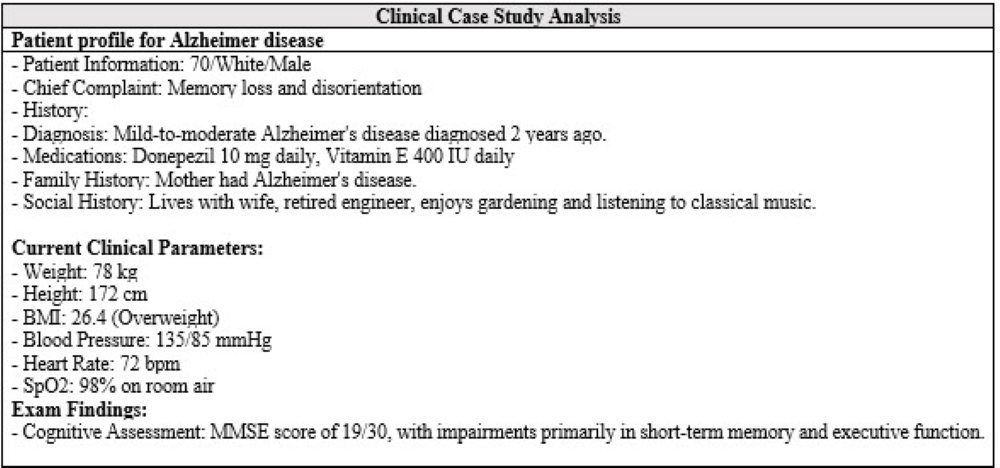}
\caption{Patient 
 Profile 1: Alzheimer's disease.} 
\label{fig:patientprofile1}
\end{figure}

\begin{figure}[H]

\includegraphics[width=\columnwidth]{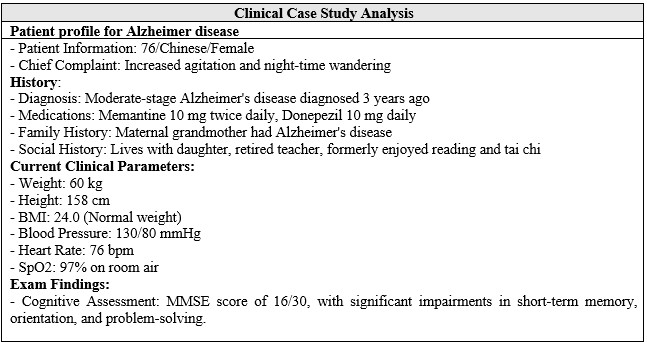}
\caption{Patient Profile 2: Alzheimer's disease.} 
\label{fig:patientprofile2}
\end{figure}

\begin{figure}[H]
\includegraphics[width=.8\columnwidth]{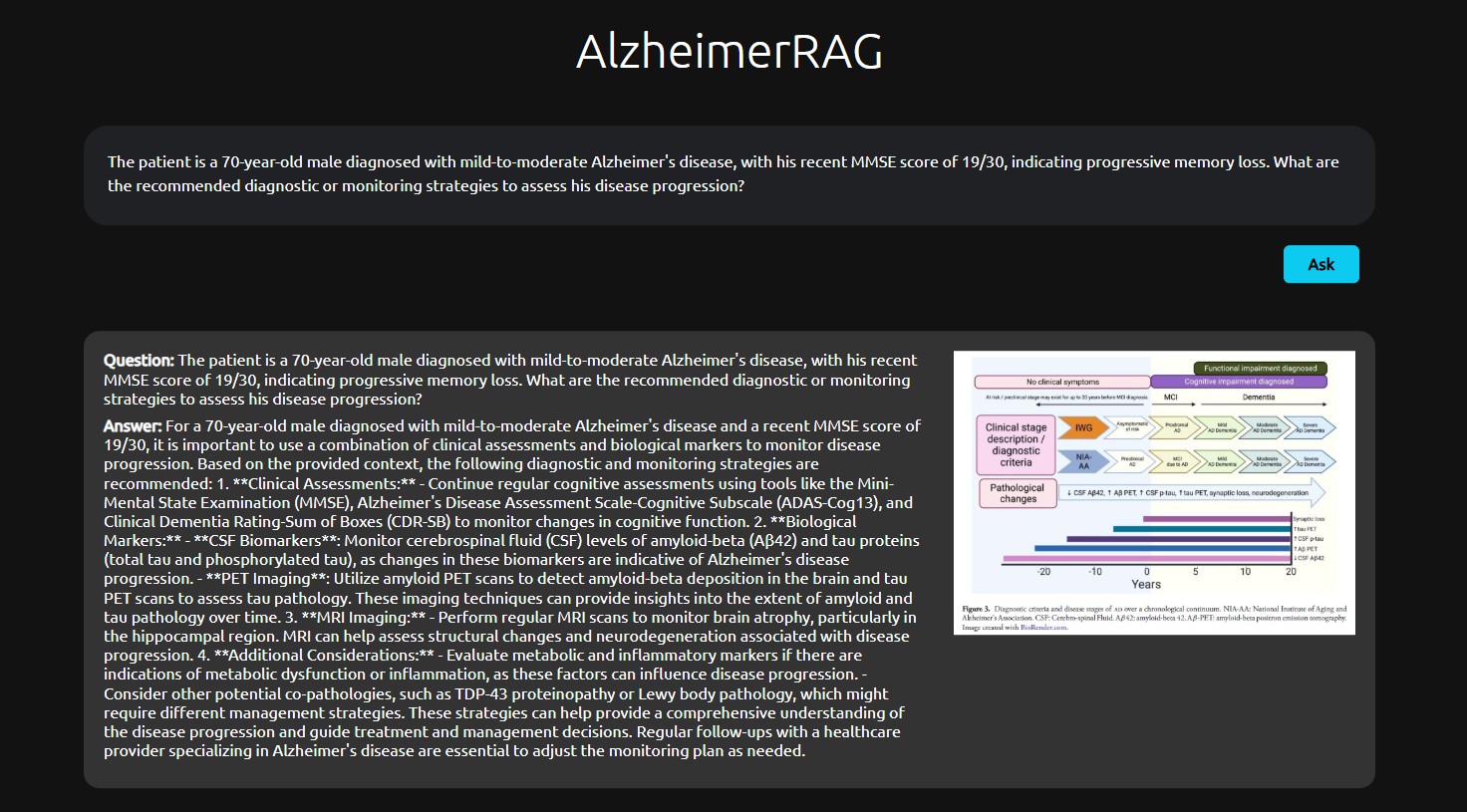}
\caption{AlzheimerRAG 
 response: Patient 1---Early Diagnosis and Monitoring.} 
\label{fig:alzheimerragresp1}
\end{figure}

\begin{figure}[H]
\includegraphics[width=.9\columnwidth]{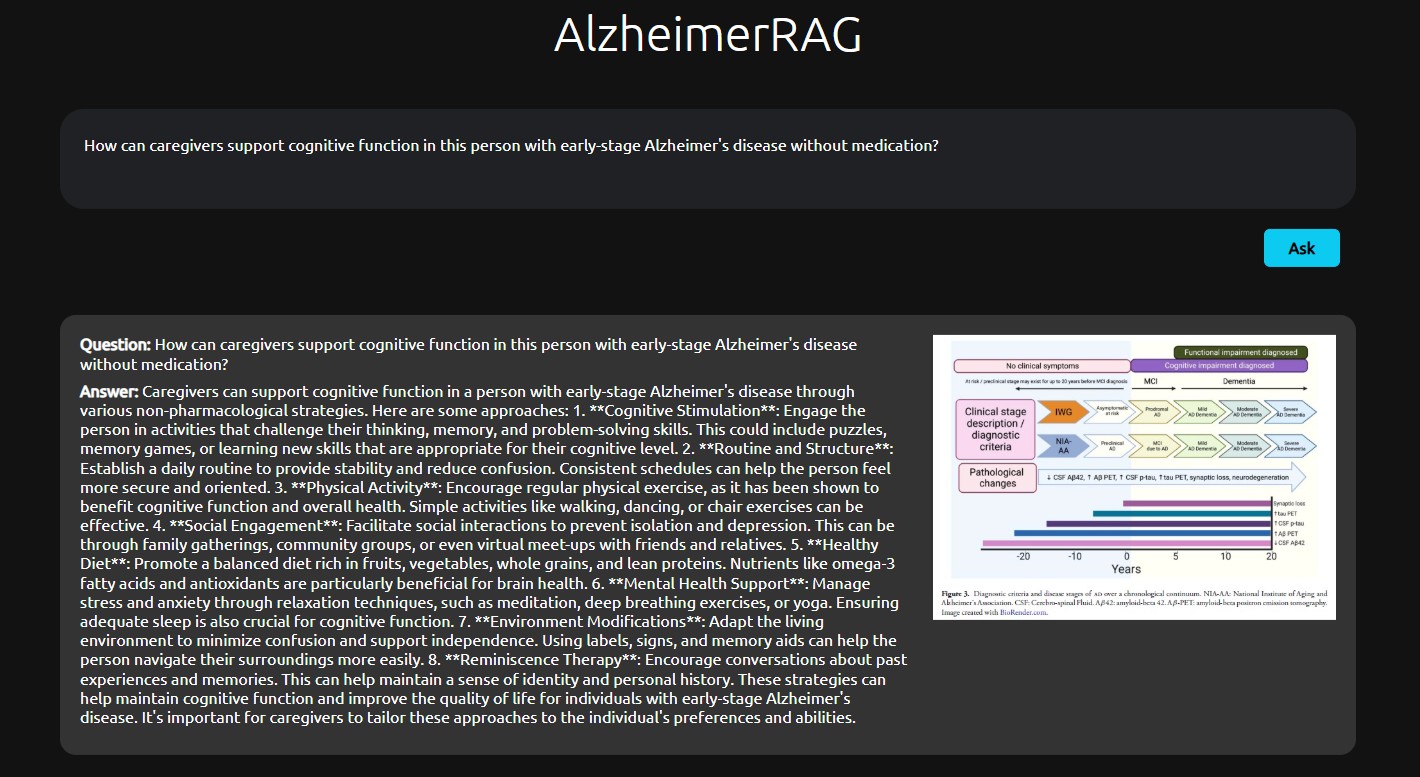}
\caption{AlzheimerRAG response: Patient 1---Non-Pharmacological Interventions.} 
\label{fig:alzheimerragresp2}
\end{figure}

\begin{figure}[H]
\includegraphics[width=.9\columnwidth]{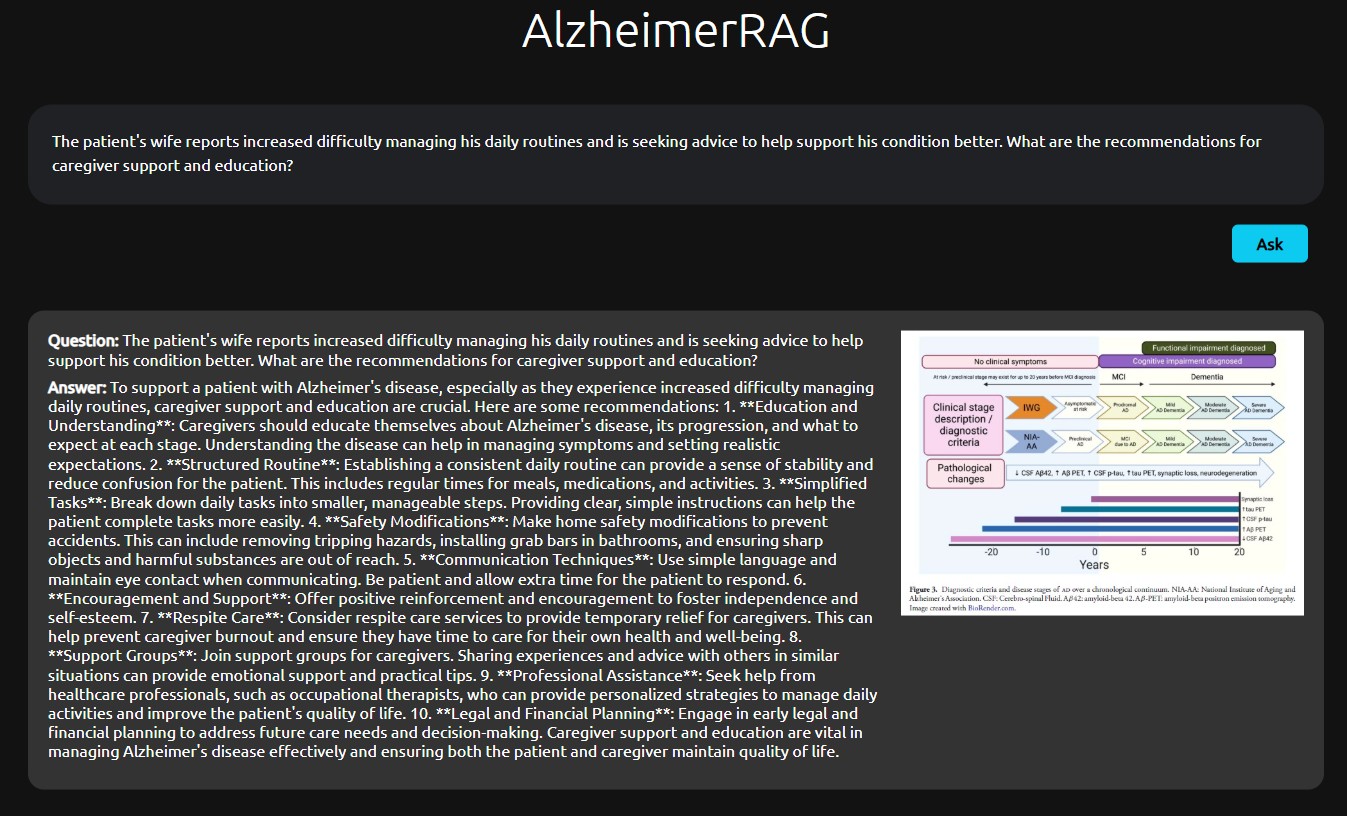}
\caption{AlzheimerRAG response: Patient 1---Caregiver Support and Education.} 
\label{fig:alzheimerragresp3}
\end{figure}

\begin{figure}[H]
\includegraphics[width=.9\columnwidth]{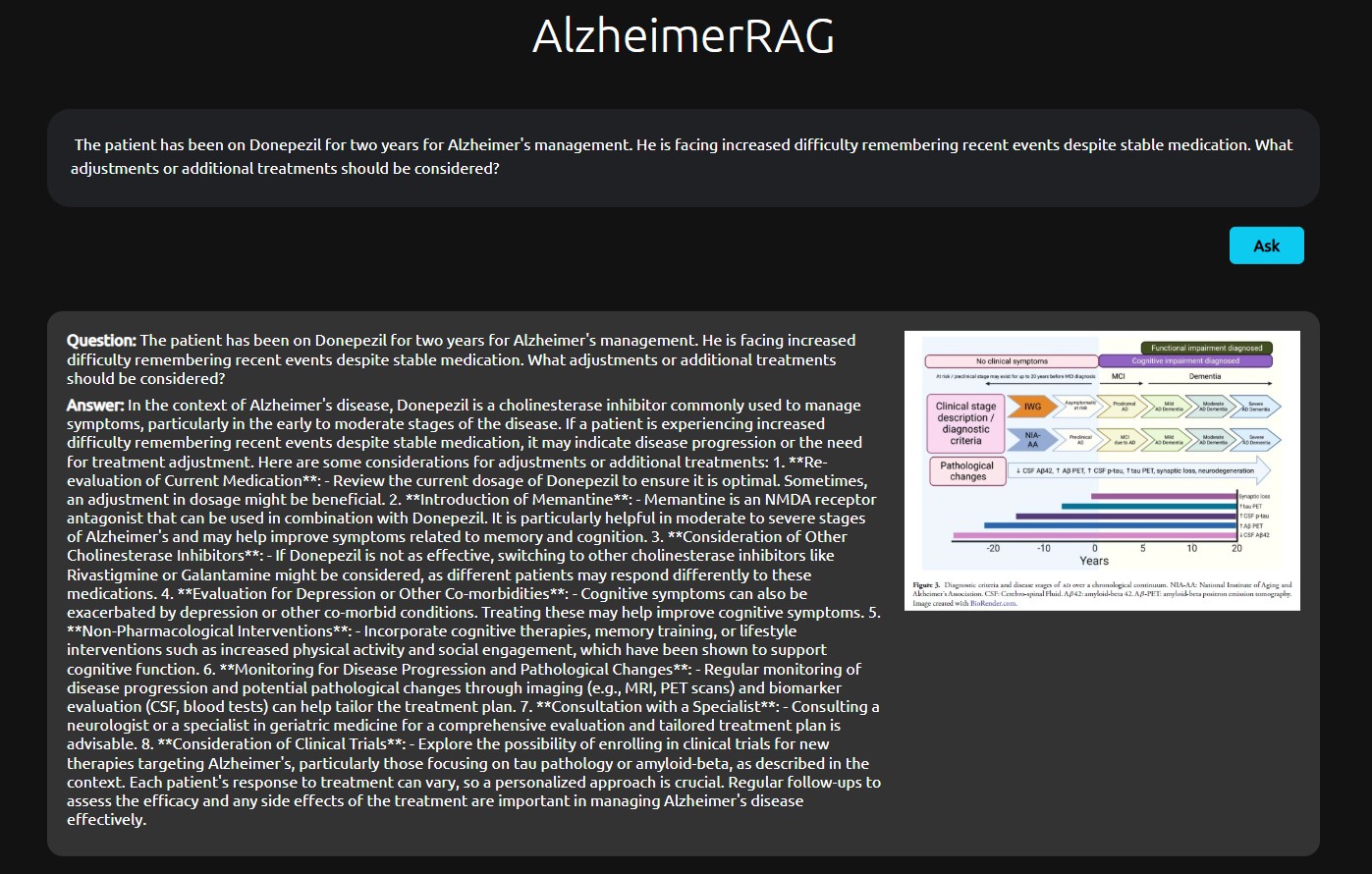}
\caption{AlzheimerRAG response: Patient 1---Medication Management.} 
\label{fig:alzheimerragresp4}
\end{figure}

\begin{figure}[H]
\includegraphics[width=.9\columnwidth]{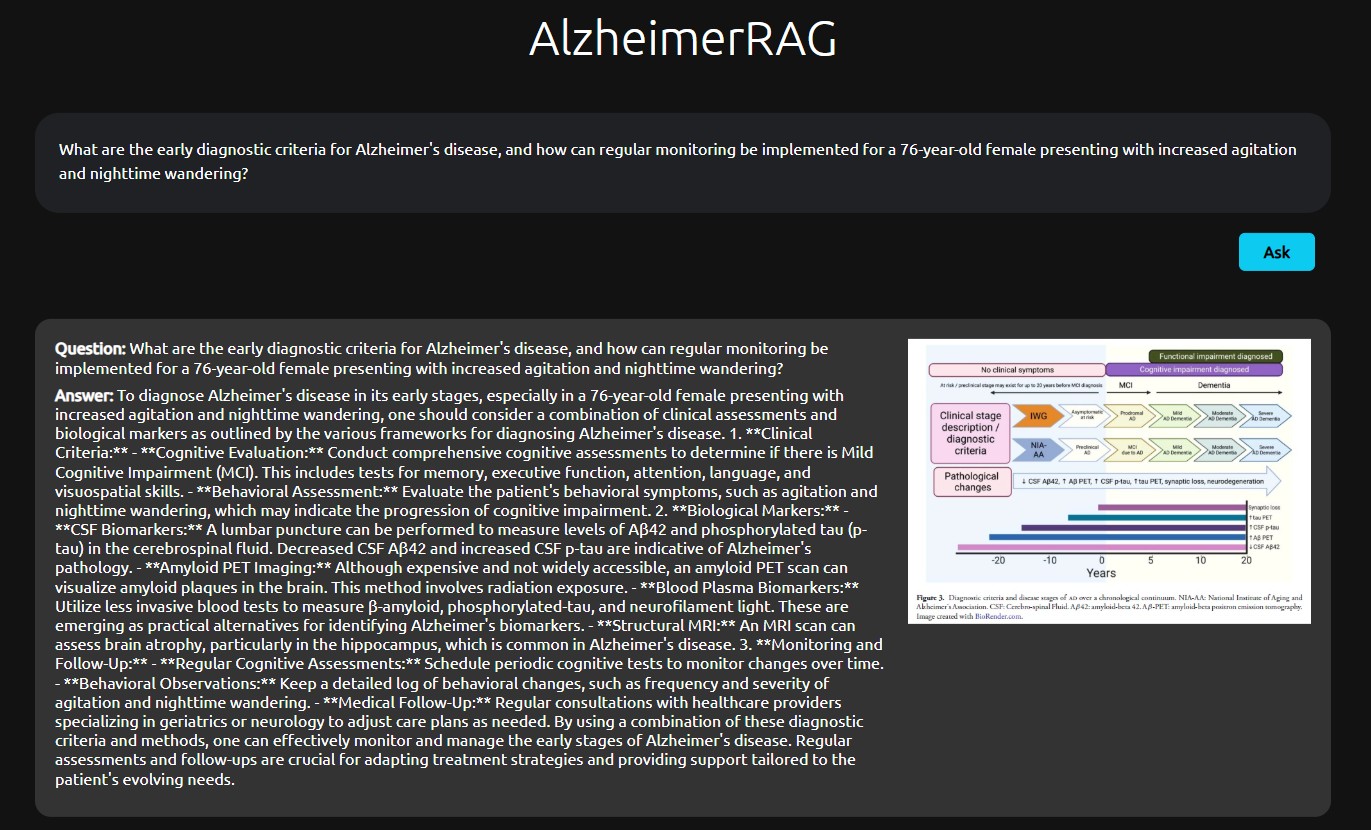}
\caption{AlzheimerRAG response: Patient 2---Early Diagnosis and Monitoring.} 
\label{fig:alzheimerragresp5}
\end{figure}

\begin{figure}[H]
\includegraphics[width=.9\columnwidth]{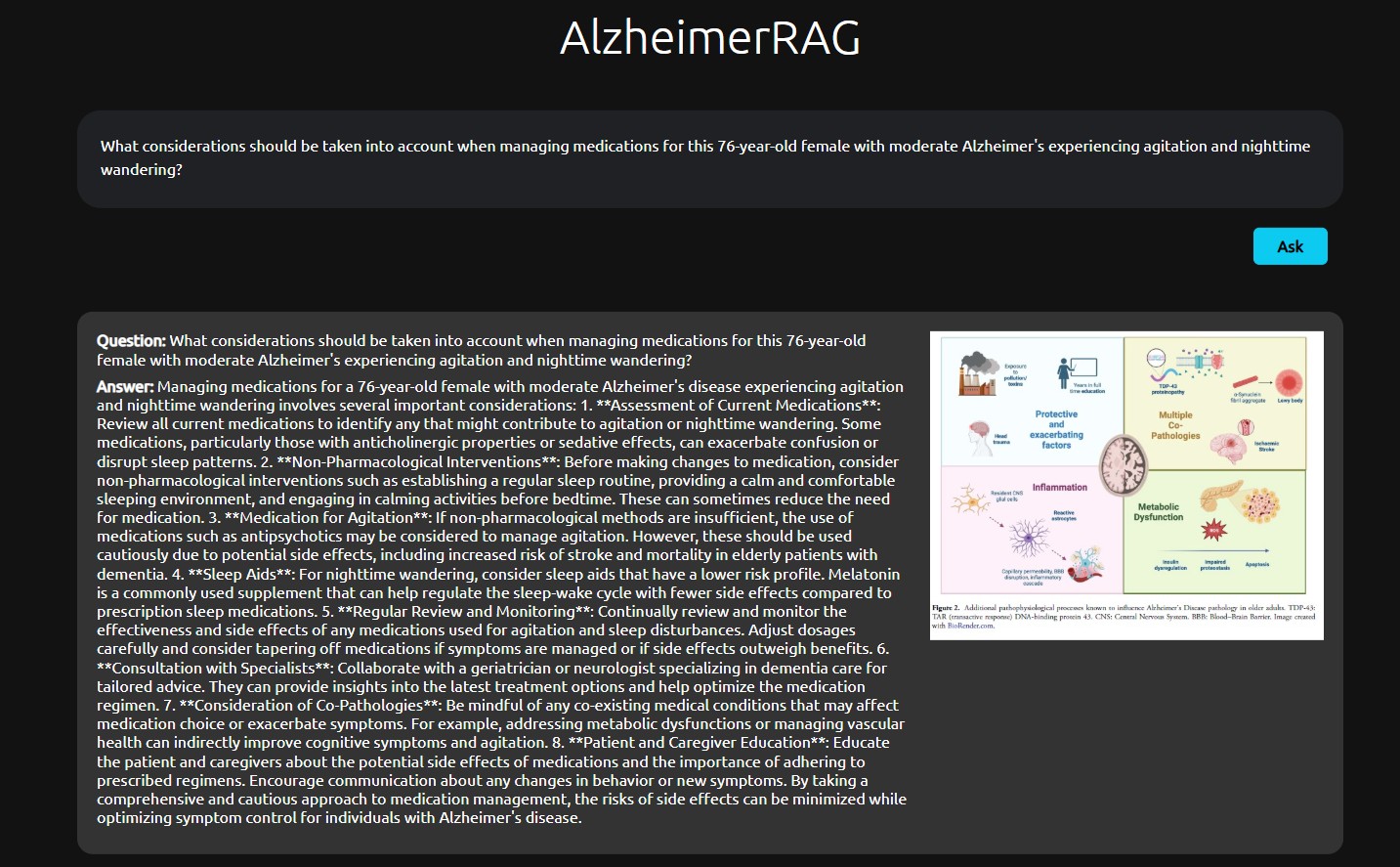}
\caption{AlzheimerRAG response: Patient 2---Medication Management.} 
\label{fig:alzheimerragresp6}
\end{figure}

\begin{figure}[H]
\includegraphics[width=.9\columnwidth]{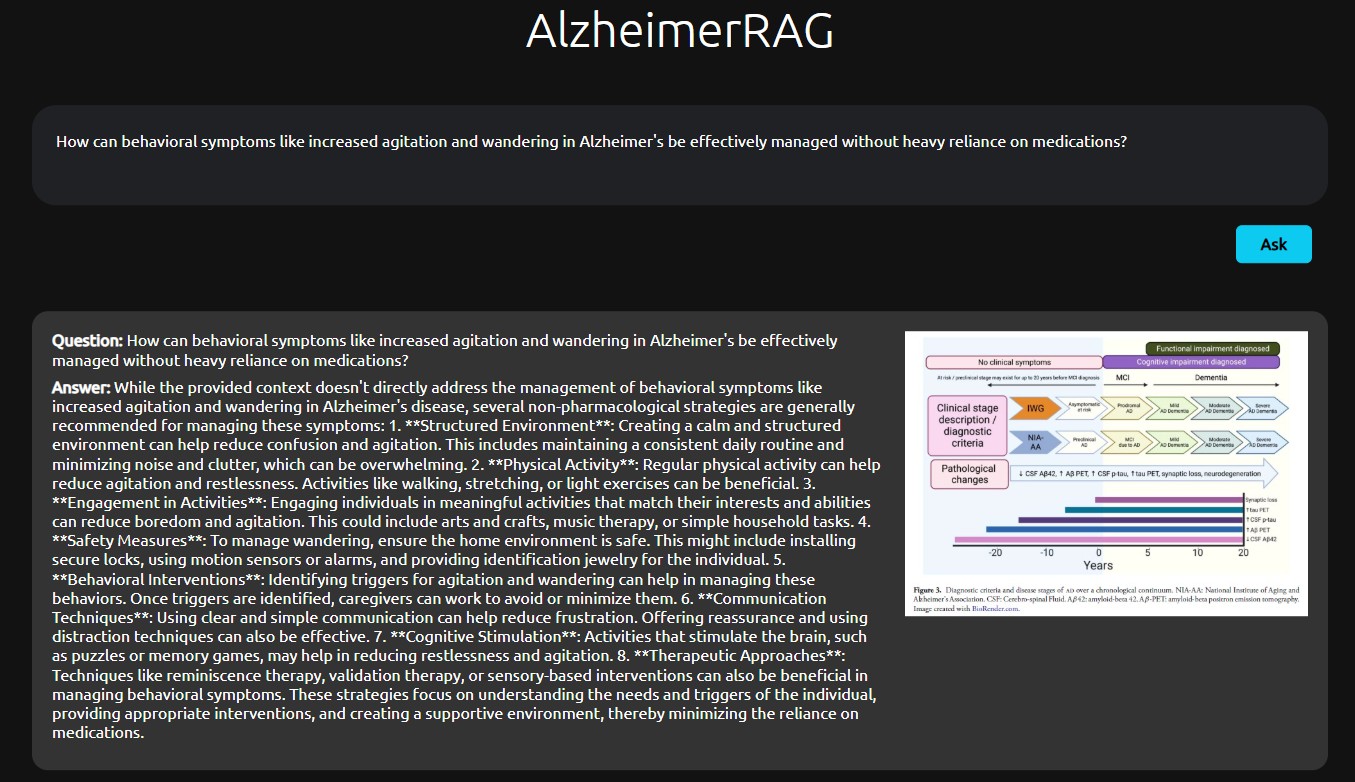}
\caption{AlzheimerRAG response: Patient 2---Behavioral Symptom Management.} 
\label{fig:alzheimerragresp7}
\end{figure}

To illustrate the AlzheimerRAG outputs with sample human-generated responses from domain experts, we provide two detailed patient profiles for Alzheimer's disease (AD) in Figures \ref{fig:patientprofile1} and  \ref{fig:patientprofile2}. We input queries with tailored questions, answered by domain experts in the AlzheimerRAG application, for each clinical scenario we designed, and retrieved the responses. The human-generated responses for queries curated for Patient Profile 1 to different clinical scenarios are presented in Boxes \ref{box2}--\ref{box5}. The human-generated responses curated for queries for Patient Profile 2 to different clinical scenarios are presented in Boxes \ref{box6}--\ref{box8}. The corresponding AlzheimerRAG responses for Patient Profile 1 are provided in Figures \ref{fig:alzheimerragresp1}--\ref{fig:alzheimerragresp4}. Similarly, for Patient Profile 2, AlzheimerRAG responses are depicted in Figures \ref{fig:alzheimerragresp5}--\ref{fig:alzheimerragresp7}.

\vspace{-9pt}


\vspace{-9pt}

\begin{boxenv}[H]
\caption{Patient 1---Clinical scenario: Early Diagnosis and Monitoring.}
\fbox{\small\parbox{\columnwidth-2\fboxsep}{
\begin{itemize}
    \item \textbf{Clinical Scenario Query}: The patient is a 70-year-old male diagnosed with mild-to-moderate Alzheimer's disease, with his recent MMSE score of 19/30, indicating progressive memory loss. What are the recommended diagnostic or monitoring strategies to assess his disease progression?
    \item \textbf{Human-Generated Response}:
    \begin{itemize}
    \item \textbf{Cognitive Testing}: Continue regular cognitive assessments such as the Montreal Cognitive Assessment (MoCA) every 6-12 months to monitor changes.
    \item \textbf{Neuroimaging}: Consider an MRI to evaluate brain atrophy and rule out other neurological conditions. 
    \item \textbf{Biomarker Analysis}: Use cerebrospinal fluid (CSF) biomarkers or PET imaging, if available, to track amyloid-beta and tau protein levels, particularly in research settings.
    \end{itemize}
\end{itemize}}} \label{box2}
\end{boxenv}

\begin{boxenv}[H]
\caption{Patient 1---Clinical scenario: Non-Pharmacological Interventions.}
\fbox{\small\parbox{\columnwidth-2\fboxsep}{
\begin{itemize}
    \item \textbf{Clinical Scenario Query}: How can caregivers support cognitive function in this person with early-stage Alzheimer's disease without medication?
    \item  \textbf{Human-Generated Response}: 
    \begin{itemize}
    \item \textbf{Cognitive Stimulation}:  Engage the patient in activities like puzzles, memory games, or learning a new hobby.
    \item  \textbf{Physical Exercise}: Encourage regular physical exercise, such as walking, to improve brain health.
    \item \textbf{Social Interaction}: Promote social engagement through group activities or community programs.
    \item  \textbf{Balance Nutrition}: Ensure a balanced diet rich in omega-3 fatty acids, antioxidants, and vitamins.
\end{itemize}
\end{itemize}}} \label{box3}
\end{boxenv}

%

\vspace{-12pt}
\begin{boxenv}[H]
\caption{Patient 1---Clinical scenario: Caregiver Support and Education.}
\fbox{\small\parbox{\columnwidth-2\fboxsep}{
\begin{itemize}
    \item   \textbf{Clinical Scenario Query}: The patient's wife reports increased difficulty managing his daily routines and is seeking advice to help support his condition better. What are the recommendations for caregiver support and education?
    \item  \textbf{Human-Generated Response}: 
    \begin{itemize}
    \item \textbf{Caregiver Education}:  Provide educational resources on Alzheimer's disease progression, specific behavioral strategies, and effective communication techniques
    \item \textbf{Support Groups}: Recommend participation in Alzheimer's support groups to share experiences and strategies with other caregivers.
    \item \textbf{Daily Routine Planning}: Develop structured routines to reduce anxiety and disorientation for both patient and caregiver, integrating familiar activities.
    \item \textbf{Respite Care}: Ensure a balanced diet rich in omega-3 fatty acids, antioxidants, and vitamins.
\end{itemize}
\end{itemize}}} \label{box4}
\end{boxenv}

\vspace{-12pt}

\begin{boxenv}[H]
\caption{Patient 1---Clinical scenario: Medication Management.}
\fbox{\small\parbox{\columnwidth-2\fboxsep}{
\begin{itemize}
    \item \textbf{Clinical Scenario Query}: The patient has been on Donepezil for two years for Alzheimer's management. He is facing increased difficulty remembering recent events despite stable medication. What adjustments or additional treatments should be considered?
    \item   \textbf{Human-Generated Response}: 
    \begin{itemize}
    \item  \textbf{Medication Review}: Ensure adherence to the prescribed Donepezil regimen and evaluate liver function to monitor for side effects.
    \item \textbf{Add Memantine}: Consider introducing Memantine, especially since symptoms are progressing, offering additional cognitive support in moderate stages.
    \item \textbf{Vitamin E and Antioxidants}: Continue Vitamin E supplementation as a supportive measure, respecting current evidence and guidelines.
\end{itemize}
\end{itemize}}} \label{box5}
\end{boxenv}

%
%

\vspace{-12pt}

\begin{boxenv}[H]
\caption{Patient 2---Clinical scenario: Early Diagnosis and Monitoring.}
\fbox{\small\parbox{\columnwidth-2\fboxsep}{
\begin{itemize}
    \item  \textbf{Clinical Scenario Query}: What are the early diagnostic criteria for Alzheimer's disease, and how can regular monitoring be implemented for a 76-year-old female presenting with increased agitation and nighttime wandering?
    \item  \textbf{Human-Generated Response}:
\begin{itemize}
    \item \textls[-15]{Use comprehensive clinical evaluations, including patient history and physical examination.}
    \item Conduct cognitive testing using tools such as the Mini-Mental State Examination (MMSE) or Montreal Cognitive Assessment (MoCA).
    \item Consider neuroimaging techniques, like MRI, to assess brain structural changes.
    \item  Implement regular monitoring every 6 to 12 months to track cognitive decline.
    \item Engage caregivers to note behavioral changes and symptoms.
    \item Utilize wearable technology to track daily activities and monitor wandering behavior.
\end{itemize}
\end{itemize}}} \label{box6}
\end{boxenv}

\vspace{-12pt}
\begin{boxenv}[H]
\caption{Patient 2---Clinical scenario: Medication Management.}
\fbox{\small\parbox{\columnwidth-2\fboxsep}{
\begin{itemize}
    \item  \textbf{Clinical Scenario Query}: What considerations should be taken into account when managing medications for this 76-year-old female with moderate Alzheimer's experiencing agitation and nighttime wandering?
    \item  \textbf{Human-Generated Response}:
\begin{itemize}
    \item  Continue cognitive enhancers like Memantine and Donepezil, adjusting doses as needed.
    \item Evaluate the need for antipsychotic medications for managing severe agitation, considering side effects.
    \item Regularly review and adjust medications based on patient response and tolerability.
    \item Monitor drug interactions, especially if the patient is on multiple medications.
    \item Explore adding new treatments only if current management is ineffective.
\end{itemize}
\end{itemize}}} \label{box7}
\end{boxenv}

%
%

\vspace{-12pt}

\begin{boxenv}[H]
\caption{Patient 2---Clinical scenario: Behavioral Symptom Management.}
\fbox{\small\parbox{\columnwidth-2\fboxsep}{
\begin{itemize}
    \item \textbf{Clinical Scenario Query}: How can behavioral symptoms like increased agitation and wandering in Alzheimer's be effectively managed without heavy reliance on medications?
    \item   \textbf{Human-Generated Response}:
\begin{itemize}
    \item Identify triggers for agitation and wandering by tracking behavioral patterns.
    \item Create a peaceful, structured environment with consistent routines.

    \item Engage patients in soothing activities, such as pet or music therapy.
    \item Redirect attention when agitation occurs, employing distraction techniques rather than confrontation.
    \item Schedule engaging activities in the late afternoon or early evening to prevent nighttime wandering.
    \item Train caregivers in behavioral management techniques to ensure uniform care strategies. 
\end{itemize}
\end{itemize}}} \label{box8}
\end{boxenv}

%

\vspace{-12pt}

\vspace{-9pt}

\vspace{-9pt}

\vspace{-9pt}

\vspace{-9pt}

\section{Conclusion and Future Work}

The AlzheimerRAG application represents a significant advancement in biomedical research, particularly in understanding and managing Alzheimer's disease. By integrating multimodal data—including textual information from PubMed articles, imaging studies, and clinical trial scenarios—this innovative retrieval-augmented generation (RAG) tool provides a comprehensive platform for analyzing complex biomedical data. The use of cross-modal attention fusion enhances the alignment and processing of diverse data types, leading to improved accuracy in generating insights relevant to diagnosis, treatment planning, and understanding the pathophysiology of Alzheimer's disease. The experimental results indicate that AlzheimerRAG outperforms existing methodologies in terms of accuracy and robustness, demonstrating the value of a multimodal approach in addressing the complexities inherent in Alzheimer's disease research. While it exhibits low hallucination rates, the risks of generating misleading information in nuanced clinical scenarios remain; as discussed in Appendix Section \ref{appendix:ethics} and \ref{appendix:hallucination}, it necessitates further research and clinical validation for real-world safety and applicability. Future enhancements could expand its scope to other neurodegenerative disorders, such as Parkinson’s, incorporate additional data sources (e.g., wearable devices, electronic health records (EHRs)), refine the user interface for improved interpretability, and optimize clinical trial support through enhanced patient recruitment and monitoring. Continuous improvements informed by user feedback will further enhance its utility and functionality for researchers and clinicians. In summary, while the AlzheimerRAG shows great promise in enhancing Alzheimer's disease research, pursuing these outlined future directions will be essential for maximizing its impact in clinical settings.



\vspace{6pt} 





\authorcontributions{Conceptualization, methodology, experiments, and the original manuscript preparation and writing were undertaken by A.K.L. The application was developed by A.K.L. and validated by Q.V.H. Review and editing of the manuscript were performed by Q.V.H. All authors have read and agreed to the published version of the manuscript.}

\funding{This research is funded by the Natural Sciences and Engineering Research Council of Canada (NSERC) research grant RGPIN/6686-2019.}

 \dataavailability{There is no new dataset developed for this research. 

 
}

\acknowledgments{We would like to thank the researchers from the Vector Institute in Toronto, Canada, for their invaluable contributions, including expert guidance on the clinical study and participation in curating the human-generated responses for the tool evaluation.}

\conflictsofinterest{ The authors declare no conflicts of interest.} 





\appendixtitles{yes} 
\appendixstart
\appendix
\section{Ethical Consideration Statement} 
\label{appendix:ethics}
AlzheimerRAG prioritizes ethical integrity by using only publicly available PubMed data, avoiding private patient information, and adhering to ethical standards without requiring Institutional Review Board approval. While rigorous filtering and cross-modal attention mitigate biases from historical PubMed imbalances, users must interpret results cautiously due to potential underrepresentation. Outputs are traceable to sources, though cross-modal complexity limits full transparency. Despite low hallucination rates (6\%), clinical oversight remains critical to validate recommendations and address outdated or conflicting data.

The system aims to accelerate Alzheimer’s research and aid underserved regions, yet risks of bias perpetuation or misinterpretation necessitate clear disclaimers and user education. Future commitments include continuous updates with the latest data, partnerships with clinicians and ethicists, and enhancements to align with real-world needs, ensuring the responsible integration of biomedical workflows while balancing societal benefits with ethical safeguards.

\section{RAG Hallucination in Medical Inference}
\label{appendix:hallucination}
While powerful, retrieval-augmented generation (RAG) systems risk generating hallucinations, factually incorrect, or unsupported outputs when synthesizing medical information. In clinical contexts, such errors could lead to harmful misdiagnoses, treatment inaccuracies, or propagation of outdated practices.

\subsection{Causes of Hallucination}
\begin{itemize}
    \item \textbf{Gaps in Retrieved Evidence:} 
 If retrieved documents lack sufficient or conflicting data, models may ``fill in'' gaps with speculative content.
    \item \textbf{Overgeneralization:} Models might conflate findings from unrelated studies or misattribute causal relationships.
    \item \textbf{Ambiguous Queries:} Poorly phrased user inputs (e.g., "Does amyloid-beta cause dementia?") may trigger oversimplified or misleading responses.
\end{itemize}

\subsection{Mitigation Strategies}
\begin{itemize}
    \item \textbf{Strict Evidence Grounding:}AlzheimerRAG restricts responses to directly cited passages from retrieved PubMed articles, minimizing unsupported claims.
    \item \textbf{Uncertainty Flagging:} The system explicitly flags low-confidence responses when retrieved data are sparse or conflicting.
    \item \textbf{Cross-Modal Verification:} Visual data are cross-referenced with textual findings to validate claims (e.g., correlating amyloid-beta plaques with cognitive decline).
    \item \textbf{Human-in-the-Loop Validation:} Clinicians review high-stakes outputs (e.g., treatment recommendations) before deployment, ensuring alignment with established guidelines.
\end{itemize}

\subsection{Recommendations for Users}
\begin{itemize}
    \item Treat AI-generated inferences as decision-support tools, not definitive medical advice.
    \item Verify critical claims against peer-reviewed guidelines (e.g., NIH Alzheimer’s).
    \item Report hallucinations via transparent feedback mechanisms to enable iterative model improvement.
\end{itemize}

\begin{adjustwidth}{-\extralength}{0cm}

\reftitle{References}





\PublishersNote{}
\end{adjustwidth}
\end{document}